\documentstyle[12pt, graphics]{article}

\newcommand{\nc}{\newcommand}
\nc{\beq}{\begin{equation}}
\nc{\eeq}{\end{equation}}
\nc{\beqa}{\begin{eqnarray}}
\nc{\eeqa}{\end{eqnarray}}

\input epsf
\newwrite\ffile\global\newcount\figno \global\figno=1

\def\writedef#1{}
\def\figin{\epsfcheck\figin}\def\figins{\epsfcheck\figins}
\def\epsfcheck{\ifx\epsfbox\UnDeFiNeD
\message{(NO epsf.tex, FIGURES WILL BE IGNORED)}
\gdef\figin##1{\vskip2in}\gdef\figins##1{\hskip.5in}% blank space instead
\else\message{(FIGURES WILL BE INCLUDED)}%
\gdef\figin##1{##1}\gdef\figins##1{##1}\fi}
\def\figinsert{}
\def\ifig#1#2#3{\xdef#1{fig.~\the\figno}
\writedef{#1\leftbracket fig.\noexpand~\the\figno}%
\figinsert\figin{\centerline{#3}}\medskip\centerline{\vbox{\baselineskip12pt
\advance\hsize by -1truein\center\footnotesize{  Fig.~\the\figno.} #2}}
\bigskip\endinsert\global\advance\figno by1}
\def\endinsert{}
\begin{document}

\title{\large{\bf On Spherically Symmetric Breathers in Scalar Theories}}

\author{
James N. Hormuzdiar\thanks{james.hormuzdiar@yale.edu} \\
Department of Physics, \\
Yale University, New Haven CT 06520 \\ \\
Stephen D.H.~Hsu\thanks{hsu@duende.uoregon.edu} \\
Department of Physics, \\
University of Oregon, Eugene OR 97403-5203 \\ \\   }

\date{May, 1999}

\maketitle

\begin{picture}(0,0)(0,0)
%\put(350,350){draft version}
\put(350,345){YCTP-P16-99}
\put(350,360){OITS-675}
\end{picture}
\vspace{-24pt}

\begin{abstract}
We develop an algorithm which can be used to 
exclude the existence of classical breathers
(periodic finite energy solutions) in scalar field theories, and 
apply it to several cases of interest.  
In particular, the technique is used to show that a pair of 
potentially periodic solutions of the 3+1 Sine-Gordon Lagrangian,
found numerically in earlier work \cite{us}, are 
not breathers.  These ``pseudo-breather states'' do have a signature in 
our method, which we suggest can be used to find similar quasi-bound state 
configurations in other theories.
We also discuss the results of our algorithm when applied to 
the 1+1 Sine-Gordon model
(which exhibits a well-known set of breathers), and $\phi ^4$ theory.

\end{abstract}

\newpage

\section {Introduction}

The leading order chiral lagrangian description of pions, 
when evaluated on an isospin-polarized
field configuration, is identical to 
the 3+1 Sine-Gordon lagrangian
\beq
{\cal L} = {1 \over 2} ~ \partial^{\mu} \phi ~ \partial_{\mu} 
\phi+ F_{\pi}^2 m^2 \cos(\phi/F_{\pi}) ~~ .
\label {SG1}
\eeq
The motivation for considering isospin-polarization of pion fields 
arises from
the possibility of producing disoriented chiral condensate (DCC) \cite{DCC}
configurations in heavy ion collisions.
In previous work \cite {us}, we discussed the numerical properties of a 
pair of spherically symmetric, long lived
classical solutions which arise from this lagrangian 
(related studies appear in \cite {SGB} and \cite {SGB2}).  
The evolution of the two configurations is nearly periodic, with energy 
localized for hundreds of oscillation cycles.  The properties 
of the lower energy 
solution are such that both the quantum corrections and higher 
derivative terms of the Chiral Lagrangian should be negligible.
Hence it is likely to correspond to a real physical pion configuration in QCD 
(pionic breather state, or PBS), which could in principle be produced
in heavy ion collisions.
Although many of its physical properties have been studied in \cite {us}, 
some theoretical questions remained unanswered, such as whether truly
periodic, finite energy solution might exist in the 3+1 Sine-Gordon model.  

In this paper we show that the two PBS solutions do not correspond to true 
breather solutions, where a breather 
is defined to be a periodic, finite energy classical solution.  
Our method can be summarized as follows. 
First, periodic solutions $\phi(r, t)$ are classified in the large $r$ 
region based on their asymptotic behavior, 
in terms of a finite number of parameters. 
Then a numerical algorithm is used to evolve the configuration using the 
equations of motion from large $r$ to $r \rightarrow 0$.
Most choices of asymptotic behavior lead to singular behavior at $r = 0$,
and hence do not correspond to finite energy solutions.
Using the criteria of finite energy, we can search the parameter
space in a systematic manner.
In section 2, 
we review the numerical properties of the two PBS solutions.  
In section 3, we 
give an overview of our algorithm.  In section 3.1, we analyze the 
free Klein-Gordon equation and
its full set of periodic solutions.  
In section 3.2, we use the solutions of section 3.1 as
asymptotic forms for the full Sine-Gordon theory to search for breather 
solutions, and present the results.
Section 4 contains a proof that interactions do not enlarge the set
of asymptotic solutions beyond those of the free Klein-Gordon equation.  
That is, there is a one-to-one mapping between asymptotic behaviors of
the free and interacting equations of motion.
In section 5 we analyze the 1+1 Sine-Gordon and $\phi ^4$ theories, finding
that the former theory contains exact breather solutions while the latter
does not.
In Section 6 we offer concluding remarks.

\section {A review of PBS properties}

\epsfysize=8.0 cm
\begin{figure}[htb]
\center{
\leavevmode
\epsfbox{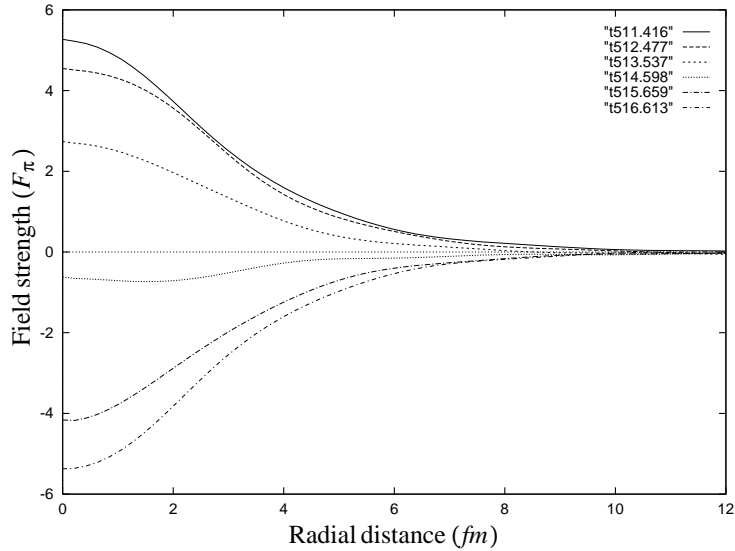}
\caption{Evolution of the lower energy PBS} 
\label{ball1}
}
\end{figure}

\epsfysize=8.0 cm
\begin{figure}[htb]
\center{
\leavevmode
\epsfbox{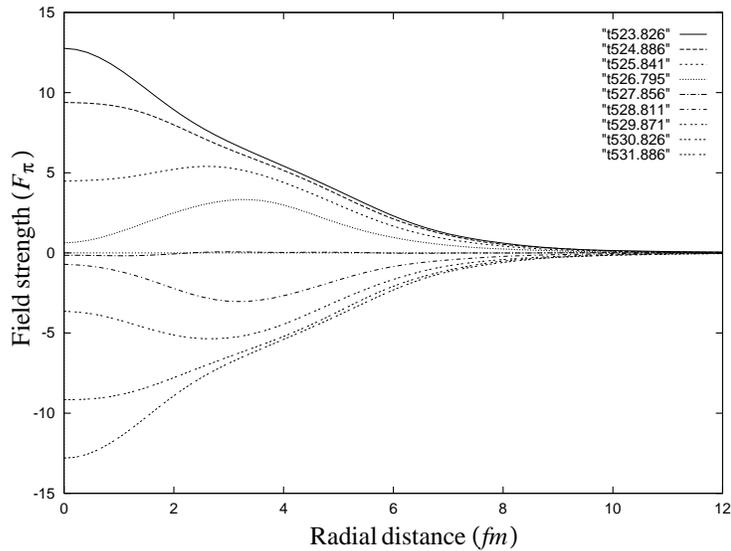}
\caption{Evolution of the higher energy PBS} 
\label{ball2}
}
\end{figure}

Both PBSs are spherically symmetric.
Imposing spherical symmetry, the equation of motion is
\beq
\ddot \phi - \phi '' - {2 \over r} \phi ' = -\sin \phi ~~ ,
\eeq
where $\dot f \equiv {d f \over d t}$ and $f' \equiv {d f \over d x}$.

Figures \ref {ball1} and \ref {ball2} show the numerically generated 
evolution of the two breather candidates, 
exhibiting the $\phi(r, t)$ vs. $r$ dependence at different times.  
The nearly periodic evolution
continues virtually unchanged for hundreds of cycles but 
eventually breaks apart, dissipating energy to 
radial infinity.

The frequency and approximate lifetime of the lower energy 
PBS are $0.1 ~ {\rm cycles /(fm/c)}$ and $850 {\rm fm/c}$.
For the higher energy PBS, they are $0.0619 ~ {\rm cycles /(fm/c) }$ and 
$4000 {\rm fm/c}$.  Additional searches have not indicated a third PBS state.

By an appropriate scaling of $\phi$ and $x^{\mu}$, the Sine-Gordon 
lagrangian (equation \ref {SG1}) can be written as
\beq
{\cal L} = {1 \over 2} ~ \partial^{\mu} \phi ~ \partial_{\mu} \phi + 
\cos \phi ~~ .
\eeq
In these units, the numerically calculated frequencies of the 
PBSs are $\omega = 0.9 \pm 0.05$ 
and $\omega = 0.58 \pm 0.02$.

\section {The Algorithm}

For finite energy solutions, 
$\phi \rightarrow 0$ as $r \rightarrow \infty$
\footnote {Sine-Gordon finite energy solutions could converge to any 
multiple of $2\pi$, but these solutions are equivalent by a 
translation in $\phi$-space to those
with $\phi \rightarrow 0$.}.
As $\phi \rightarrow 0$, nonlinearities in the equation of 
motion become negligible, hence
the $r \rightarrow \infty$ behavior of any breather solution must 
approach a periodic solution to the noninteracting Klein-Gordon theory.

The strategy of the following algorithm is as follows. First, we
solve for the most general periodic, finite energy
solutions of the Klein-Gordon equation.  These solutions are 
used as large r asymptotic forms and are evolved inward to the
origin using the 
full Sine-Gordon equation of motion.  
Only solutions which are finite and differentiable at the origin can 
correspond to true breather states.
Since the space of asymptotic forms is parameterized by a 
finite number of variables, 
a comprehensive search can be initiated which in the end can 
rule out the existence of breathers within a certain frequency range.

\subsection {Asymptotic Forms}

It can be shown that the most general periodic, spherically symmetric 
classical solutions 
to the free Klein-Gordon equation are linear combinations of 
\beq
\phi_{\omega}(r, t) ~=~ {1 \over r} \sin(k r - \theta_k) \sin(\omega t - \theta_{\omega}) ~~ ,
\label {generalperiodicform}
\eeq
where 
\beq
w^2 = k^2 + m^2 ~~ .
\eeq
If $m = 0$, these solutions have infinite energy for all $\omega$, and 
no breathers exist.  
For nonzero $m$, the set of solutions for 
$0 < \omega < m$ have imaginary k, corresponding to the 
finite energy solutions of the form
\footnote {These are also not breathers, because 
$\phi'_{\omega}(0, t) \neq 0$, and hence there are
no true breathers of the free Klein-Gordon equation.}
\beq
\tilde \phi_{\omega}(r, t) ~=~ {1 \over r} \exp(-\tilde k r) 
\sin(\omega t - \theta) ~~ ,
\eeq
where
\beq
\omega ^2 + \tilde k^2 = m^2 ~~ .
\eeq
(Since these are the only types of solutions we are interested in, 
we will drop the tilde from now on).
The most general finite energy periodic solution is therefore
\beq
\phi ~=~ {1 \over r} \sum_{j = 1}^{k_j ~ {\rm real}} a_j \exp(-k_j r) 
\sin(j \omega_f t - \theta_j) ~~,
\eeq
where $\omega_f$ is the fundamental frequency in the Fourier series, 
and $k_j = \sqrt {m^2 - (\omega j)^2}$.
Note that in the case that $0.5 m < \omega_f < m$, this sum has only one term
\beq
\phi ~=~ {1 \over r} ~ a_1 \exp(-k_1 r) \sin(\omega_f t) ~~,
\label {asympt-form}
\eeq
hence the full set of solutions is parameterized by two variables, 
$a_1$ and $\omega_f$.  The frequencies of
the two numerical PBS solutions fall well within this frequency range.

In the remainder of this section we will work in units with $m=1$.

\subsection {The Search for Breathers}

Using the asymptotic form in equation \ref {asympt-form} as 
$r \rightarrow \infty$ boundary conditions, 
the Sine-Gordon equations of motion can be integrated inward to 
$r = 0$ (see appendix A for computational details).  
In general, solutions obtained in this way 
are neither finite nor differentiable 
at the origin, and therefore 
do not correspond to physical breather states.  
For small r, the numerically obtained solutions can be fit to the form
\beq
{A(t) \over r} + B(t) ~~ ,
\eeq
and the integral
\beq
S \equiv \int_0^{2 \pi / \omega_f} d t ~ |A(t)|
\eeq
can be used as a measure of how singular each solution is at the 
origin\footnote {Although $S$ is 
obtained by fitting to a 1/r form, a nonzero value of $S$ can be shown to 
also indicate singularities of the type $\phi \sim 1/r^n$ with $n>1$.  
A first derivative fit
near the origin would be even more powerful 
(capable of fitting derivative and logarithmic singularities),
but tends to yield ever increasing values of S as the fit is done 
closer to the origin, which is less manageable. 
It will therefore be avoided until cases where it is necessary, 
such as the 1+1 Sine-Gordon equation 
discussed later in this paper.}.  
A similar measure of differentiability could be defined 
(for instance by fitting 
to a form $A(t) ~ r + b(t)$), but as will soon be shown, in 
the present case not even the singularity condition is fulfilled,
so it is not necessary.

The size of the singularity $S$ is calculated for various 
values of $a_1$ and $\omega$.  To fill $a_1-\omega$ space as 
uniformly as possible
values are randomly chosen in the range $0.5 < \omega_f < 1$ and $a_1 > 0$.
In practice an upper limit on $a_1$ has to be set, so a value
is chosen which is much larger than that of the two known 
pseudo-breather states.
Also, the lower limit of $a_1$ had to be raised slightly (from 0 to 1.2), 
because 
$a_1 \rightarrow 0$ corresponds to the trivial solution $\phi(r, t) = 0$, 
which lowers
the value of $S$ to zero across all values of $\omega$.

\begin{figure}[htb]
\center{
\leavevmode
\epsfbox{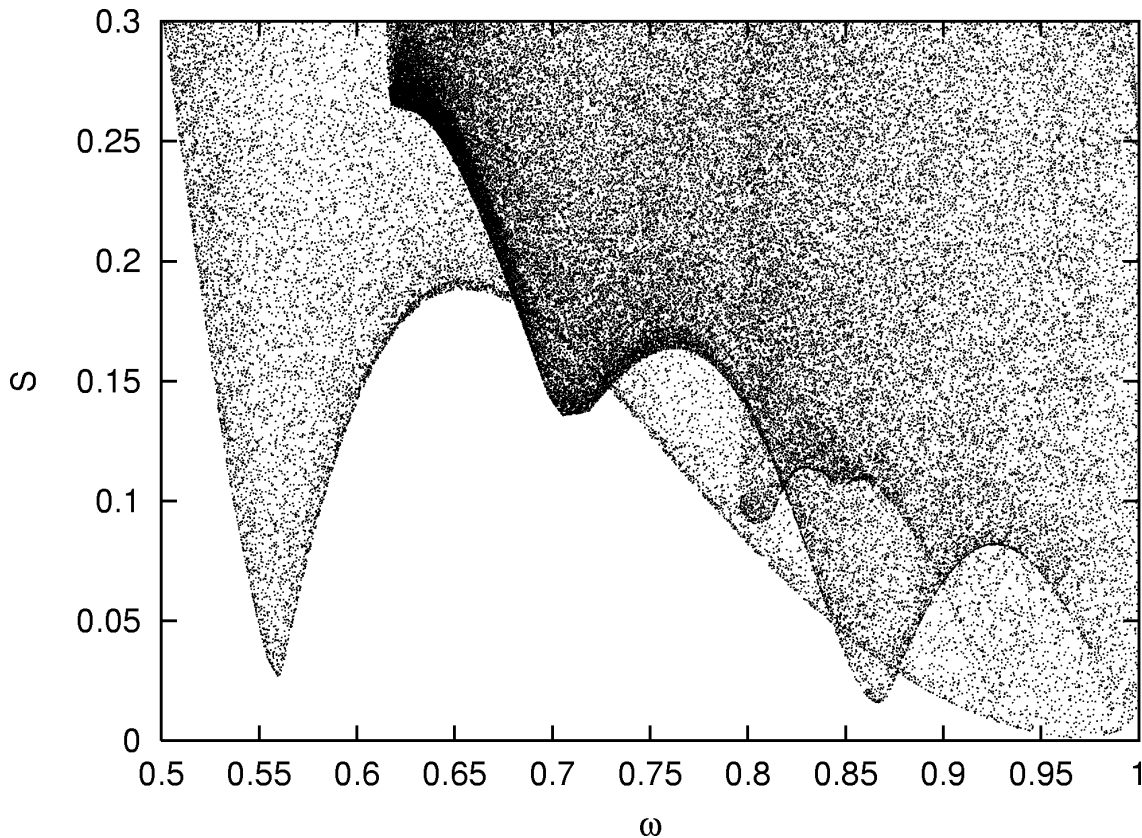}
\caption{The singularity $S$ vs. oscillation frequency $\omega_f$} 
\label{Svsomega}
}
\end{figure}

Figure~\ref{Svsomega} shows a plot of $S$ vs. $\omega_f$ 
(the value of $a_1$ is not shown, so the figure
can be thought of as a transparent projection plot).  
As can be seen, this plot has 
three prominent minima.  Two occur at the frequencies of the previously
discovered PBSs, $\omega \approx 0.560$ and $\omega \approx 0.865$.  
Neither minima goes all the way to zero,
and hence they do not represent exact breathers.
The third minima at $\omega = 1$ corresponds to the trivial solution
\beq
\phi(r, t) = A \sin(m t - \theta) ~~ ,
\eeq
with infinitesimal $A$.
                     
\epsfysize=11.0 cm

\begin{figure}[htb]
\center{
\leavevmode
\epsfbox{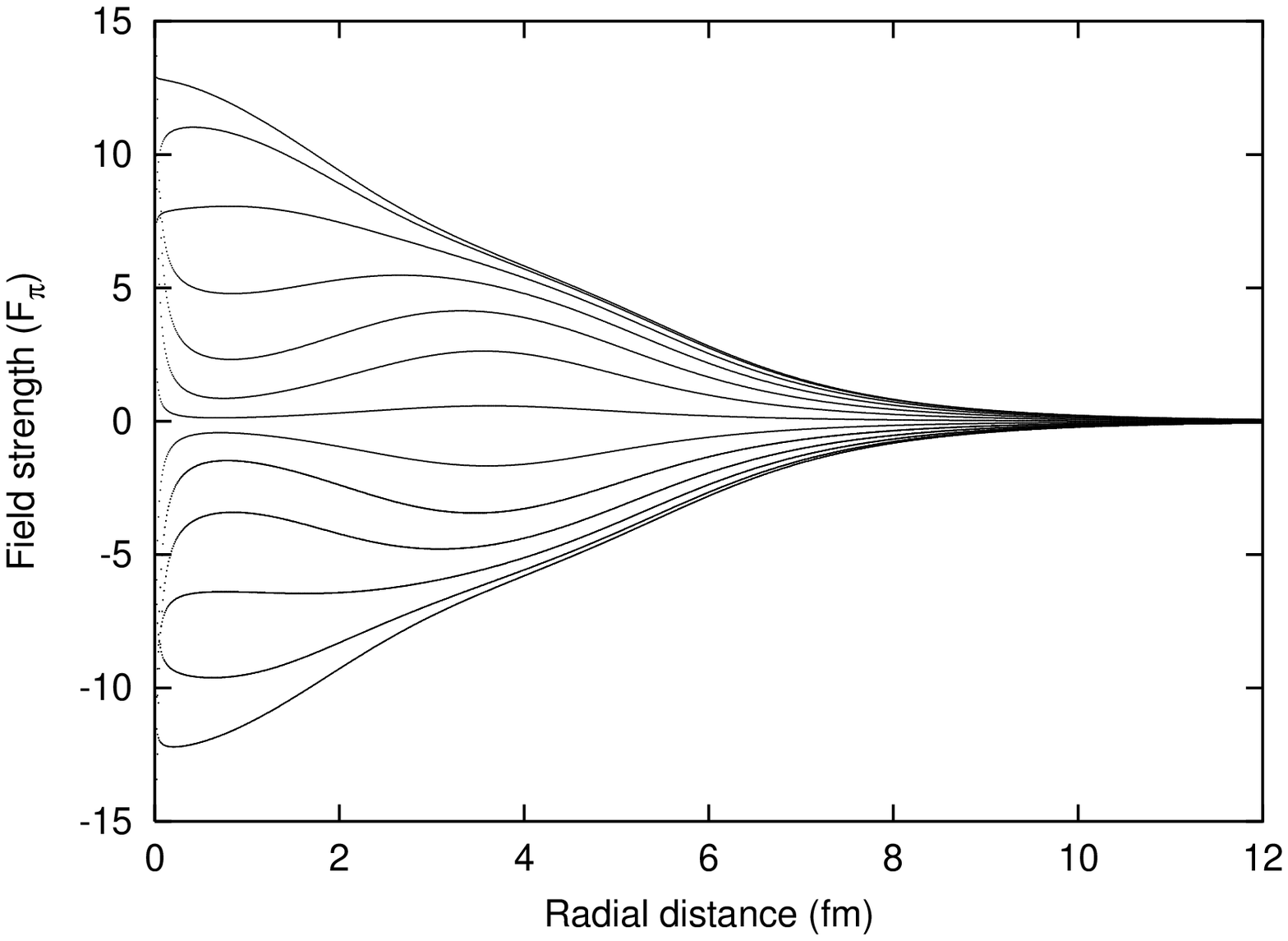}
\caption{Evolution of $\phi(r, t)$ at the $\omega \approx 0.560$ 
minima of figure \ref {Svsomega}} 
\label{dball2}
}
\end{figure}

As further evidence supporting our hypothesis that these 
minima correspond 
to the previously discovered PBS states, figure \ref {dball2} shows the 
shape of a $\phi(r, t)$ calculated near the minima at 
$\omega \approx 0.560$. Notice the similarity with figure \ref {ball2}, with 
the exception of $r \approx 0$ where figure \ref {dball2} becomes singular.

\epsfysize=8.0 cm

We now define the solutions corresponding to the minima of the $S$ vs. 
$\omega_f$ curve 
as 'pseudo-breather' states (a useful reinterpretation of the acronym 'PBS', 
which previously
referred to 'Pionic Breather State').  Although not true breathers, 
pseudo-breathers can often
exhibit very interesting behavior.  As was discussed in \cite {us}, the 
two PBS configurations arise readily from multi-domain DCC configurations, 
have characteristic energies and frequencies, and might extend the 
lifetime of classical pion
configurations by more than an order of magnitude.  

%New insights into non-linear 
%equations of motion can 
%therefore be obtained by generating the $S$ vs. $\omega$ plot.

\section {A More Formal Justification}

At this point questions still remain about the comprehensive nature of the 
solutions 
obtained in the previous section.  It is unclear how the addition of 
interactions might enlarge the full set of periodic solutions.  
Perhaps each asymptotic form corresponds
to multiple solutions, and integrating inward only uncovers one.  
In this section we reanalyze the
equations of motion by expanding $\phi$ in a power series.  
Without solving the resulting complicated set 
of equations, we study its general nature to answer the question just 
presented, and therefore
verify that the previously described algorithm is valid.

In section 4.1, we work out a general form of the solutions.  
In section 4.2, we use this general form
to solve the free Klein-Gordon equations as a test case for the full theory, 
which we finish with in section 4.3.

\subsection {A Series Form for the Solutions}

First we show that the most general periodic solution with frequency 
$0.5 < \omega < 1$ is of the form \footnote {In many particular cases such as the
Sine-Gordon or $\phi ^4$ lagrangians, the sum in this general form starts at
$j_1 = 3$ and $j_2 = 3$.  This is however a particular case of the form as given, 
and therefore need not be considered separately.}
\beq
\phi(r, t) = a ~ \sin(\omega t) ~ {e^{-kr} \over r} +
\sum_{j_1 = 2}^\infty \sum_{j_2 = 2}^\infty A_{j_1, j_2}(t) ~ 
{e^{-j_2 k r} \over r^{j_1}} ~~ ,
\label {generalform}
\eeq
where the $A_{j_1, j_2}(t)$ are unspecified functions, and 
$k^2 + \omega ^2 = m^2$.

Parameterizing the interaction strength as $\lambda$, 
\beq
{\cal L} = {1 \over 2} \partial^{\mu} \phi \partial_{\mu} 
\phi - {1 \over 2} m^2 \phi^2 + \lambda ~ V(\phi) ~~ ,
\eeq
we can now write the solution $\phi$ as a series in $\lambda$, 
\beq
\phi(r, t) = \sum_{n = 0}^{\infty} \lambda^n \phi_n(r, t) ~~ .
\label {lambdaseries}
\eeq
We will inductively show that each term is of the form of equation 
\ref {generalform}

Substituting \ref {lambdaseries} into the equations of motion, yields
\beq
[\partial^2 + m^2] ~ \phi_n(t, r)~=~
{\cal P}_n[\phi_0(r, t), ~\phi_1(r, t), .... ~ \phi_{n-1}(r, t)] ~~ ,
\label {lambdaserieseq}
\eeq
where ${\cal P}_n$ are polynomials expressions related to the 
Klein-Gordon interaction term in a complicated way.
The equation for $n = 0$ is just the Klein-Gordon equation of motion
\beq
[\partial^2 + m^2] ~ \phi_0(t, r)~=~ 0 ~~ ,
\eeq
hence its solution is the usual one given by equation \ref {asympt-form},
\beq
\phi ~=~ a ~ \sin(\omega t) ~ {e^{-k r} \over r} ~~,
\eeq
This is also the first term in equation \ref {generalform}.

Iteratively substituting in the expressions for 
$\phi_0, \phi_1, .... \phi_{n-1}$, we can solve for $\phi_n$.

To show that higher
$\phi_n$'s are in the form of the remaining term in equation 
\ref {generalform}, first note 
that since $\cal P$ is polynomial whose lowest order term is at 
least quadratic, inserting 
$\phi_0, \phi_1, .... \phi_{n-1}$, yields something of the form
\beq
\phi(r, t) = \sum_{j_1 = 2}^\infty \sum_{j_2 = 2}^\infty 
A_{j_1, j_2}(t) ~ {e^{-j_2 k r} \over r^{j_1}} ~~ ,
\label {secondform}
\eeq
(this differs from equation \ref {generalform} only in that it is 
missing the first term).
Equation \ref {lambdaserieseq} is a second order linear 
inhomogeneous differential equation whose most general
solution is
\beq
\phi_n(r, t) = a ~ \sin(\omega t) ~ {e^{-kr} \over r}  + Q(r, t) ~~ ,
\label {solnforphi}
\eeq
where $a$ is an undetermined variable and Q(r, t) is any particular solution.
It can be shown that $[\partial^2 + m^2] ~ Q$ spans the space given by 
equation \ref {secondform}
if $Q$ is of the form in equation \ref {generalform}.  A $Q$ in the 
desired form can therefore be chosen to satisfy
the differential equation, and by equation \ref {solnforphi}, 
$\phi_n$ is also in this form.

\subsection {The Klein-Gordon Case}

Before we work with the full equation of motion, we will use the 
form of equation \ref {generalform} in
the free Klein-Gordon Lagrangian to derive the full solution for a given 
large $r$ behavior of $\phi$ (in the form of equation \ref {asympt-form}), 
and show that this is unique.  Although
this result is trivial, it sets up a formalism which can be naturally 
extended for the 
next section when we will repeat this for the full theory 
including interactions.

Substituting the form of equation \ref {generalform} into the 
Klein-Gordon equation of motion yields the
decoupled set of equations
\beq
\ddot A_{2, j_2}(t) + [-j_2^2 ~ k^2 + m^2] A_{2, j_2}(t) = 0 ~~ ,
\label {diff-eq-A1}
\eeq
\beq
\ddot A_{3, j_2}(t) + [-j_2^2 ~ k^2 + m^2] A_{3, j_2}(t) = 
4 j_2 k A_{2, j_2}(t) ~~ ,
\label {diff-eq-A2}
\eeq
and then for $j_1 > 3$,
\beq
\ddot A_{j_1, j_2}(t) + [-j_2^2 ~ k^2 + m^2] A_{j_1, j_2}(t) 
= 2 j_2 k (j_1 - 1) A_{j_1 - 1, j_2}(t) 
+ j_1 (j_1 - 1) A_{j_1 - 2, j_2}(t) ~~ .
\label {diff-eq-A3} 
\eeq
These are each second order linear inhomogeneous differential equations 
with general solution
\beq
A_{j_1, j_2}(t) = a_{j_1, j_2} \sin(\omega_{j_2} t - \theta_{j_1, j_2}) 
+ P_{j_1, j_2}(t)~~ ,
\label {formforA}
\eeq
where $a_{j_1 j_2}$ and $\theta_{j_1, j_2}$ are undetermined constants, 
$P_{j_1, j_2}(t)$ is any particular 
solution, and
\beq
\omega_{j_2}^2 + (k j_2)^2 = m^2 ~~ .
\label {condition}
\eeq
Note however that each $a_{j_1, j_2}$ is forced to be zero.  
The $\sin(\omega_{j_2} t - \theta_{j_1, j_2})$
is an on-resonance driving term for the next differential equation (with
$j_1$ incremented by one), 
and the associated solution is not periodic.  All freedom in the choice of 
$a_{j_1, j_2}$ is lost.  The driving terms 
for each differential equation iteratively become zero and therefore each 
$P_{j_1, j_2}(t) = 0$.

The most general solution with the given lowest order asymptotic 
term is therefore 
\beq
\phi ~=~ {1 \over r} ~ a ~ \exp(-k r) \sin(\omega t) ~~ .
\eeq

\subsection {Adding Interactions}

Now we repeat the analysis in the last section for the case where there is 
an interaction term present,
and arrive at our final conclusion for this section, that the solution 
to the full equation (interaction
present) is completely determined by the frequency and size of the first order 
asymptotic piece that is a solution to 
the free Klein-Gordon theory.

Repeating the same analysis with an interaction term changes 
equation \ref {diff-eq-A1}-\ref {diff-eq-A3} to 
\beq
\ddot A_{2, j_2}(t) + [-j_2^2 ~ k^2 + m^2] A_{2, j_2}(t) = 
{\cal P}_{1, j_2}[A_{1, 1}, A_{2, 2}, ....] ~~ ,
\label {full-A-diff-eq1}
\eeq
\beq
\ddot A_{3, j_2}(t) + [-j_2^2 ~ k^2 + m^2] A_{3, j_2}(t) = 
4 j_2 k A_{2, j_2}(t) +
{\cal P}_{2, j_2}[A_{1, 1}, A_{2, 2}, ....] ~~ ,
\label {full-A-diff-eq2}
\eeq
and then for $j_1 > 3$,
\begin {eqnarray}
\ddot A_{j_1, j_2}(t) + [-j_2^2 ~ k^2 + m^2] A_{j_1, j_2}(t) 
= 2 j_2 k (j_1 - 1) A_{j_1 - 1, j_2}(t) 
+ j_1 (j_1 - 1) A_{j_1 - 2, j_2}(t) + 
\\ \nonumber
{\cal P}_{j_1, j_2}[A_{1, 1}, A_{2, 2}, ....] ~~ .
\label {full-A-diff-eq3}
\end {eqnarray}

The actual form of ${\cal P}_{j_1, j_2}$ is related to the 
Klein-Gordon interaction term in a 
complicated way, but two things are clear.  First, since the interaction 
contains no 
linear terms, ${\cal P}_{j_1, j_2}$ is at least quadratic in the 
$A_{j_1, j_2}$'s, and can not
explicitly cancel the linear terms already present from 
equations \ref {diff-eq-A1}-\ref {diff-eq-A3}.  
Second, each ${\cal P}_{j_1, j_2}$ only contains
terms $A_{\tilde j_1, \tilde j_2}(t)$ with $\tilde j_1 < j_1$ 
and $\tilde j_2 < j_2$.  This 
allows us to solve iteratively beginning with the lower order solutions, 
yielding once again simple second order 
linear inhomogeneous differential equations.

Unlike in the noninteracting case, after inserting in the lower order 
solutions $A_{j_1, j_2}(t)$, ${\cal P}_{j_1, j_2}$ may
have a term in it proportional to the on-resonance driving piece,
$\sin(\omega_{j_2} t - \theta_{j_1, j_2})$.  This doesn't add 
any freedom in our choices of $a_{j_1, j_2}$
however, in that now these values must be chosen to cancel the 
resonant pieces.
This constraint completely specifies the form of each $A_{j_1, j_2}(t)$, hence 
there is no additional freedom in the choice of solutions.

The final solution is therefore unique given its lowest order 
large $r$ asymptotic behavior.

\section {Other Examples}

\epsfysize=8.0 cm
\begin{figure}[htb]
\center{
\leavevmode
\epsfbox{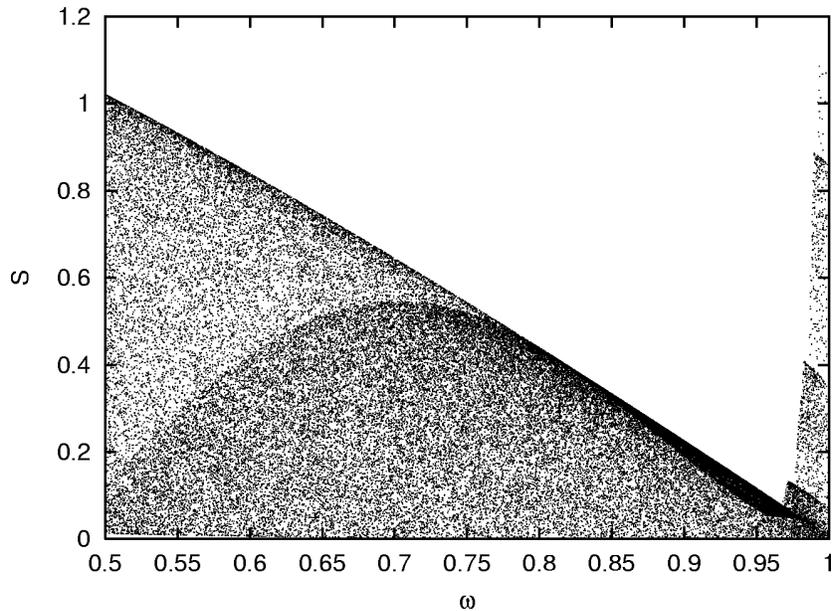}
\caption{S vs. $\omega$ for the 1+1 Sine-Gordon Lagrangian} 
\label{SG11}
}
\end{figure}

\epsfysize=8.0 cm
\begin{figure}[htb]
\center{
\leavevmode
\epsfbox{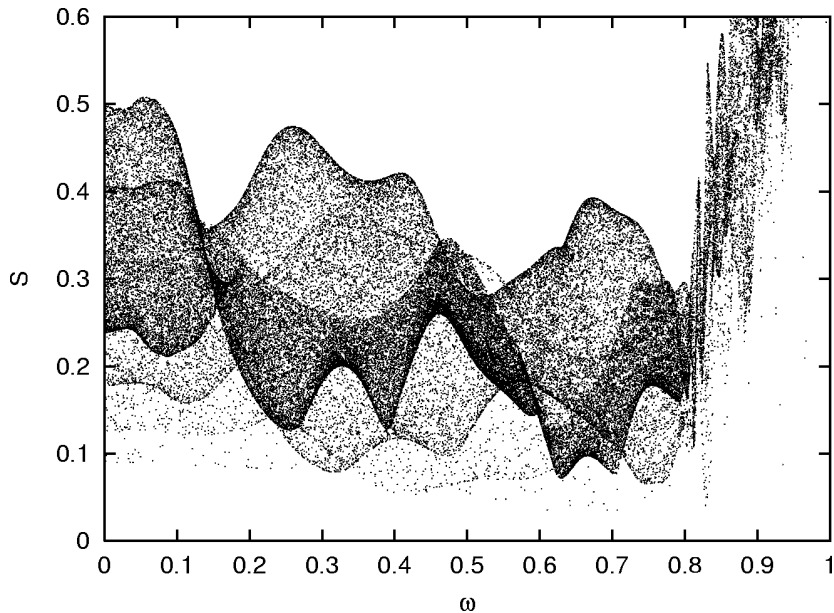}
\caption{S vs. $\omega$ for the 3+1 $\phi ^4$ theory} 
\label{phi4}
}
\end{figure}

We end by applying our method on two other examples, the 1+1 
Sine-Gordon equation, and
$\phi ^4$ theory.  The 1+1 Sine-Gordon Lagrangian
is a well studied model with known breathers, and is thus used as a 
test of our algorithm.

Minor modifications had to be made to work in 1+1 dimensions.  
The asymptotic form in 
equation \ref {asympt-form} was changed to 
\beq
\phi ~=~ a_1 \exp(-k_1 r) \sin(\omega_f t) ~~.
\eeq
Also, in 1+1 dimensions, $\phi(r, t)$ does not readily become 
singular at $r = 0$, hence we tested the 
differentiability of $\phi$ at zero.

The $S$ vs. $\omega$ plots for each Lagrangian appear in figures 
\ref {SG11} and \ref {phi4}.  In the 
1+1 Sine-Gordon plot the values of $S$ go to zero across all 
values of $\omega$, indicating
a class of breathers as expected.  The $\phi ^4$ plot stays 
positive for all values of $\omega$,
hence in 3+1 dimensional $\phi ^4$ theory, there are no 
breathers in the frequency range $0.5 < \omega < 1$.

\section {Conclusions}

In this paper we have developed a numerical search algorithm for
spherically symmetric breathers in spin zero theories.  
This algorithm was used to show that
breather-like states found in a previous work were not actual breathers, 
but rather a new type of object, a 'pseudo-breather', which, while
long-lived, is ultimately unstable. We showed that the asymptotic
behavior of periodic solutions
which might have finite energy can be
parameterized in terms of a finite number of parameters. 
In the case of interest, $0.5m < \omega_f < m$, there are only two
parameters: frequency $\omega_f$ and magnitude $a_1$. 
An extensive search within the Sine-Gordon model failed
to uncover any values 
of $\omega_f, a_1$ for which the solution is non-singular at the origin.
The results for this model are shown in figure \ref {Svsomega}.

We have, as of yet, left the region $0 < \omega_f < 0.5m$ unexplored 
($\omega_f > m$ is automatically excluded
by large $r$ energy considerations).  It is straightforward in principal 
to extend 
the algorithm to arbitrarily small $\omega$.  This is done by 
generalizing the form of the 
asymptotic solutions.  For instance, in the region $0.25m < \omega_f < 0.5m$, 
the asymptotic form becomes
\beq
\phi ~=~ {1 \over r} ~ a_1 \exp(k_1 r) \sin(\omega_f t) 
+ {1 \over r} ~ a_2 \exp(k_2 r) \sin(2 \omega_f t - \theta_2) ~~ .
\eeq
There is a numerical trade-off, in that to cover smaller regions of 
$\omega_f$ the number 
of parameters to describe the asymptotic solution increases, eventually to 
infinity as $\omega_f \rightarrow 0$.

The condition of spherical symmetry can be relaxed slightly by adding 
higher partial wave terms
\beq
\phi(\vec x, t) = \phi_0^0 (r, t) ~ Y_0^0(\theta, \phi) ~+~ 
\phi_1^{-1} (r, t) ~ Y_1^{-1}(\theta, \phi) ~+~ \phi_1^0 (r, t) ~ 
Y_1^0(\theta, \phi) ~+~
\phi_1^1 (r, t) ~ Y_1^1(\theta, \phi) ~+~ ....
\eeq
Inserting this into the equation of motion yields a 
coupled set of equations
\beq
\ddot \phi_l^m - {\phi_l^m}'' - {2 \over r} ~ {\phi_l^m}' + { l (l+1) \over r^2} \phi_l^m 
= -V' \left( \sum_{l, m} Y_l^m \phi_l^m \right) ~~ .
\eeq
Again, the number of variables needed to parameterize 
the space of asymptotic forms increases, thus demanding more 
computational power for each
spherical harmonic added.

Although not discussed here, our algorithm can be modified to work with 
higher spin theories.
The condition of spherical symmetry greatly restricts the asymptotic forms, so 
in many cases not much more computational power is needed.

Two shortcomings of the analysis in this paper are as follows.  
First, we can not guarantee that our upper cutoff in $a_1$ hasn't 
hidden a true breather, 
and second, it is possible (as often with a numerical analysis) that some 
exceptional behavior is concealed in the gaps 
between calculated values in the $S$ vs. $\omega_f$ graph.
Due to the large range of points and small spacing in figure \ref {Svsomega}, 
this seems unlikely. However, it prevents the result from being completely 
rigorous.

\bigskip
\noindent 
The authors would like to thank R. Beals, C. Buragohain, Alan Chodos, M. G\"{o}k\c{c}eda\u{g}, Michael Ibrahim,
and Vincent Moncrief for useful discussions and comments.
This work was supported in part under DOE contracts DE-AC02-ERU3075 
and DE-FG06-85ER40224.

\bigskip

\section {Appendix A : Details of Numerical Evolution}
The spherically symmetric 3+1 Sine-Gordon equation of motion is
\beq
\ddot \phi(r, t) - \phi ''(r, t) - {2 \over r} \phi'(r, t) = - \sin(\phi) ~~ .
\eeq
Discretizing, and using the approximations
\beq
{d f(x_i) \over d x} \approx {f(x_{i+1}) - f(x_{i-1}) \over 2 \Delta x}
\eeq
and
\beq
{d ^2 f \over d x ^2} \approx {f(x_{i+1})- 2 f(x_{i}) +f(x_{i-1}) \over 
\Delta x^2} ~~ ,
\eeq
the equations of motion can be written
\beq
\phi(r_{i-1}, t_j) = {1 \over i-1}\left[ 2 \phi(r_i, t_j) ~ i - \phi(r_{i+1}, 
t_j) (i+1) 
+ \Delta x^2 ~ i ~ f \right] ~~ ,
\eeq
where
\beq
f = {\phi(r_i, t_{j+1})+\phi(r_i, t_{j-1}) - 2 \phi(r_i, t_j) 
\over \Delta t^2} + \sin(\phi (r_i, t_j)) ~~.
\eeq

Spatial boundary conditions are fit to an asymptotic form at 
$r \rightarrow \infty$.  In practice this 
boundary fit is done at a large but finite value $r = r_0$.  
Boundary conditions in the $t$ direction 
are of course periodic.
The values $r_0$, $\Delta x$ and $\Delta t$ were varied until 
the resulting evolution (and hence the
shape of figure \ref {Svsomega}) converged.

%\newpage

\end{document}